\documentstyle[11pt]{article}

\newtheorem{thm}{Theorem}[section]

\newtheorem{prop}[thm]{Proposition}

\begin{document}

\title{\bf{ Conformal Solutions of   Static Plane Symmetric Cosmological Models in  Cases
of a Perfect Fluid and  a Cosmic String Cloud }}

\author{ {\bf{Ragab M. Gad}}$^{1}$ \thanks{%
E-mail: ragab.ali@mu.edu.eg}\,\,,  {\bf{Awatif Al-Jedani1}}$^{2}$\thanks{%
E-mail: amaljedani@uj.edu.sa }  \,\, and {\bf{Shahad T. Alsulami1}}$^{2}$\thanks{%
E-mail:: sulamishahad@gmail.com } \\
\newline
{\it $^1$ Department of Mathematics, Faculty of Science, Minia University,}\\
 {\it   61915 El-Minia,  Egypt}\\
{\it $^2$
 Mathematics Department, Faculty of Science, University of
 Jeddah,}\\
 {\it  21589 Jeddah, KSA}
}

\date{\small{}}

\maketitle

\begin{abstract}
In this work, we  obtained exact solutions of Einstein's field
equations for plane symmetric cosmological models by assuming that
thy admit conformal motion. The space-time geometry of these
solutions is found to be nonsingular, non-vacuum and conformally
flat. We have shown that in the case of a perfect fluid, these
solutions have an energy-momentum tensor possessing dark energy with
negative pressure and the energy equation of state is $\rho +p=0$.
We have shown that a fluid has acceleration, rotation, shear-free,
vanishing expansion, and rotation. In the case of a cosmic string
cloud, we found that the tension density and particle density
decrease as the fluid moves along the direction of the strings, then
vanish at infinity. We shown that the exact conformal solution for a
static plane symmetric model  reduces to the the well-known anti-De
Sitter space time. We obtained that the space-time under
consideration admits a conformal vector field orthogonal to the
four-velocity vector and does not admits  a vector parallel to the
four-velocity vector. Some physical and kinematic properties of the
resulting models are also discussed.

\end{abstract}

\setcounter{equation}{0}
\section{Introduction}
General relativity, a gravitational field theory, is described from
the viewpoint of geometry and physics by Einstein's field equations,
which are highly nonlinear. Because of this nonlinearity, it becomes
very difficult to solve these equations unless we assume certain
constraints, such as symmetries, on the spacetime metric. However,
finding exact solutions to such equations and their physical
interpretations is sometimes more difficult. Despite these
difficulties, there are many exact solutions to these equations. In
addition to the exact solution, there are also non-exact solutions
that describe certain physical systems.
\par
One of the most successful ways of finding exact solutions to
Einstein's field equations has been to consider that the space-time
under study admits one of symmetries. Symmetries also provide us
with deeper insights into the properties of space-time. Besides the
interest of these symmetries from the geometric and physical aspects
of space-time, they play an important role in simplifying Einstein's
field equations and providing a classification
of space-time according to the structure of the corresponding Lie algebra.\\
In the theory of general relativity and its equivalent theories,
symmetries are studied on the basis of the geometry corresponding to
each theory. In the context of  Riemannian geometry, different types
of symmetries such as isometric, homothetic motion, conformal
motion, matter collineations, and Ricci collineations ... etc, have
been extensively studied in the theory of general relativity.\\
Of course, the most studied type of these symmetries in general
relativity is the Killing symmetry, and many examples and uses of it
are known \cite{SKMHH03}.  killing Symmetry is a special case of
homothetic symmetry whose generalization is conformal symmetry.
Duggal and Sharma \cite{DS99} presented the characterizations and
classifications of the space-times of general relativity admitting
Killing, homothetic and conformal symmetries. A more detailed
discussion of the different types of homothetic symmetry
can be found in \cite{HW78} - \cite{CC05}.\\
In a series of papers \cite{GadAlofi}-\cite{GSH21}  Gad, Alofi and
Al Mazrooei studied the homothetic symmetry using Lyra's geometry.
in this case, space-times were classified according to admit of such
symmetry. It turns out that in the case of a zero displacement
vector field, the results obtained in the context of Lyra's geometry
agree with those obtained previously in the theory of general
relativity, using Riemannian's geometry. While in the case of a
constant displacement vector field, it is not possible to compare
the results obtained in the context of Lyra's geometry with those
obtained in general relativity, using Riemannian's geometry. This
showed that in Lyra's geometry, if the displacement vector field is
taken to be constant, this does not give meaningful results. Killing
and homothetic  symmetries have also been studied in the theory of
teleparallel gravity, using Weitzenb\"{o}ck's geometry
\cite{SM09}-\cite{SA08}. In the context of Finsler's geometry,
Sanjay  et al  \cite{SNNM24} investigated the charged gravastars
with conformal motion. They examined charged gravitationally vacuum
stars under the background of Finslerian gravity with the use of the
conformal Killing vector and have considered charged stellar objects
with three different regions and distinct equation of state
parameters to
analyze the structure of such objects.\\
 In this work, our focus will be on studying the conformal symmetry of
 a static plane
symmetric cosmological model and finding exact solutions to the
Einstein field equations without assuming any restrictions either on
the variables or on the physical properties of a given space-time,
as is common in the literature. We will only assume that the model
under study admits a conformal motion.\\
One of the important properties of  conformal symmetries is they
preserve the causal character of space-time. That is, If there is a
conformal vector field in a space-time which, if the metric is
dragged along it, the causal structure of space-time remains
constant.
 One small drawback of these symmetries lies in the fact
that, unlike isometry and homothetic symmetries,  do not leave the
Einstein tensor constant, and in this respect, they can be
considered non-natural or accidental. However, although this
drawback some solutions with conformal symmetries are known.\\
Many researchers have studied  Spherically symmetric perfect-fluid
space-times admitting a conformal Killing vector field. Gad
\cite{G02} studied these solutions and  derived a different
coordinate representation of the solutions obtained in \cite{HP85}
and \cite{K94}. Exact solutions of Einstein's field equations are
found in the case when the conformal Killing vector field is
parallel to the 4-velocity vector field $u^a$. Recently, non-static
of these space-times have studied in \cite{HDO22}. Several families
of exact analytical solutions are found for different choices of the
conformal vector field in both the dissipative and the adiabatic
regime. Recent and old literature also provides some important
results using conformal Killing vector field (see for example the
references \cite{SY09} - \cite{KHBK15}).\\
The contents of the paper are organized as follows: In the next
Section, the physical and geometric parameters for  a static plane
symmetric space-time are given. We solve conformal  equations for
this space-time and get the conformal Killing vector, conformal
factor and the relation between the coefficients of the metric. The
results obtained must be satisfied Einstein's field equations, and
this is what we will do in the two subsequent  sections of Section
2. In section 3, we consider the matter represented by a perfect
fluid. In section 4, we consider the case of a cosmic strings could.
In section 5, We discus that the conformal vector field orthogonal
to the four-velocity vector and does not admits  a vector parallel
to the four-velocity vector. In section 5 we study a vector field
that is orthogonal to the four-velocity vector and a vector that is
parallel to the four-velocity vector. We discuss which vector fields
are allowed by the space-time under study and which are not.
 In section 6, Some physical and kinematic properties of the resulting
 models are also discussed. Finally, in Section 7, concluding remarks are given.

\setcounter{equation}{0}
\section{\bf{ Version of model and conformal vector field}}
Let   $M$ be a four-dimensional, Hausdorff, smooth manifold  with a
non-degenerate metric tensor  $g$
with signature $(+, -, -, -)$.\\
A  vector field $\zeta$ on  a space-time $(M,g)$ is said to be
conformal vector if the following is satisfied
\begin{equation}\label{CE}
\pounds_\zeta g_{ab}=\zeta_{a;b}+\zeta_{b;a}=2\psi(t,x,y,z)
g_{ab}\qquad \Leftrightarrow \quad \zeta_{a;b}=\psi g_{ab}+F_{ab},
\end{equation}
 where the conformal factor $\psi=\psi(x^a)$ is a scalar function , $\pounds_\zeta$
 denotes a Lie derivative operator relative to $\zeta$ and
semi-colon ($;$) denotes a covariant derivative w. r. t. the metric
connection. If $\psi_{,a}=0$, that is, $\psi$ is constant on $M$,
then the conformal vector field $\zeta$ is called homothetic (proper
homothetic vector field if $\psi=const.\neq 0$ on $M$). If $\psi =0$
on $M$,  $\zeta$  is called a Killing vector field. In components
form, the first equation in (\ref{CE}) takes the following form
\begin{equation}\label{CE2}
g_{ab,c}\zeta^c+g_{ac}\zeta^c_{,b}+g_{cb}\zeta^c_{,a}=2\psi g_{ab}.
\end{equation}
A general plane symmetric space-time can always be written
 in the following form
$$
ds^2=e^{2A(t,x)}dt^2-e^{2C(t,x)}dx^2-e^{2B(x)}(dy^2+ dz^2)
$$
For a static plane symmetric cosmological models, the coefficients
of the metric will be independent of time $t$. In this case $x$ can
be redefined to get rid of the coefficient of $dx^2$,  in the above
equation, which now reduces to the standard representation and is
given by \cite{SKMHH03}
\begin{equation}\label{met}
ds^2=e^{2A(x)}dt^2-dx^2-e^{2B(x)}(dy^2+ dz^2),
\end{equation}
with the convention $x^0=t$ (cosmic time), $x^1=x$, $x^2=y$ and
$x^3=z$  and the scale factors
$A(x)$ and $B(x)$ are functions of $x$ only.\\
As shown in \cite{SKMHH03}, the above   space-time (\ref{met})
admits four independent Killing vector fields which are as follows:
$$
\frac{\partial}{\partial t}, \frac{\partial}{\partial y},
\frac{\partial}{\partial z},
$$
$$
y\frac{\partial}{\partial z}-z\frac{\partial}{\partial y}.
$$
The physical and geometric parameters of the space-time (\ref{met})
are determined by the following: \cite{R79}
\begin{enumerate}
\item The only non-vanishing component of the 4-acceleration vector,
$\dot{u}_a  =  u_{a;b}u^b$, is
\begin{equation}
\dot{u}_1  = - A^\prime,
\end{equation}
\item  The only non-vanishing components of the rotation, $\omega_{ab}  =
u_{[a;b]}+\dot{u}_{[a}u_{b]}$, are
\begin{equation}
\omega_{01}  =-\omega_{10}=A^\prime e^{A},
\end{equation}
where prime denotes to the derivative w. r. t. $x$.
\item The expansion scalar
\begin{equation}
\Theta =   u^a_{;a} = 0 ,
\end{equation}
\item The shear scalar
$$
\sigma^2  = \sigma_{ab}\sigma^{ab},
$$
 where
$$
\sigma_{ab} =u_{(a;b)}+\dot{u}_{(a}u_{b)}
-\frac{1}{3}\Theta(g_{ab}+u_au_b).
$$
For the space-time (\ref{met}), we get
\begin{equation}
\sigma_{00}= \sigma_{11}=\sigma_{22}=\sigma_{33}=0, \quad
\sigma^2=0.
\end{equation}
That is, the space-time (\ref{met}) is shear free.
\end{enumerate}
\par
Study of conformal vector fields, $\zeta=\zeta^a(t,x, y,
z)_{a=0}^3$, on  a static plane symmetric model (\ref{met}) is based
on solving the ten reduced equations (due to the symmetry of the
metric $g_{ab}$) obtained from the first equation of (\ref{CE}). For
the space-time (\ref{met}), the  corresponding conformal equations
are given by the following system of equations:
\begin{equation}\label{1}
\zeta^0_{,0}+A^\prime \zeta^1 =\psi(x),
\end{equation}
\begin{equation}\label{2}
e^{2A}\zeta^0_{,1}-\zeta^1_{,0}=0,
\end{equation}
\begin{equation}\label{3}
e^{2A}\zeta^0_{,2}-e^{2B}\zeta^2_{,0}=0,
\end{equation}
\begin{equation}\label{4}
e^{2A}\zeta^0_{,3}-e^{2B} \zeta^3_{,0}=0,
\end{equation}
\begin{equation}\label{5}
\zeta^1_{,1} =\psi(x),
\end{equation}
\begin{equation}\label{6}
e^{2B}\zeta^2_{,1}+ \zeta^1_{,2}=0,
\end{equation}
\begin{equation}\label{7}
e^{2B}\zeta^3_{,1}+ \zeta^1_{,3}=0,
\end{equation}
\begin{equation}\label{8}
\zeta^2_{,2} +B^\prime\zeta^1=\psi(x),
\end{equation}
\begin{equation}\label{9}
\zeta^2_{,3}+ \zeta^3_{,2}=0,
\end{equation}
\begin{equation}\label{10}
\zeta^3_{,3} +B^\prime\zeta^1=\psi(x).
\end{equation}
where the commas denote partial derivatives w. r. t. the coordinate
indicated.\\
Integrating  (\ref{5}) w. r. t. $x$, we get
\begin{equation}\label{11}
\zeta^1 =\int{\psi(x)}dx +F^1(t,y,z).
\end{equation}
Using this result back into equations (\ref{1}) - (\ref{10}), taking
into account that $\psi=\psi(x)$,  and after some algebraic
calculations, the above system of equations gives the following
components of the conformal vector field and constraint relation
$$
\zeta^0=const.=c_0
$$
\begin{equation}\label{zeta1}
\zeta^1 =\frac{\psi}{B^\prime},
\end{equation}
$$
\zeta^2=const. =c_2,
$$
$$
\zeta^3=const. =c_3,
$$
$$
F^1(t,y,z)=const.=c_1
$$

\begin{equation}
A(x) =B(x).
\end{equation}
where $c_a, a= 0,1,2,3$ are  constants of integration.\\
Inserting the above results back into equation (\ref{11}) and
integrating the obtained results, we get
\begin{equation}
\psi=A^\prime e^{A}=B^\prime e^{B}.
\end{equation}
After the previous discussion, the following result can be
established.
\begin{thm}\label{thm}
A plane symmetric space-time described by the metric ansatz
(\ref{met}) admits a conformal Killing vector if  the following
conditions are satisfied
$$
A=B
$$
$$
\psi=A^\prime e^{A}.
$$
The conformal Killing vector corresponding to this case takes the
following form
\begin{equation} \label{zeta^a}
{\bf{\zeta}}=c_0\frac{\partial}{\partial t}+e^A
\frac{\partial}{\partial x} + c_2\frac{\partial}{\partial y}
+c_3\frac{\partial}{\partial z}.
\end{equation}
\end{thm}
According to the previous theory, in order to a static plane
symmetric model (\ref{met})  admits conformal motion, it must be the
well-known anti-De Sitter space-time, which takes the usual form
\begin{equation}
ds^2=e^{2A(x)}(dt^2-dy^2- dz^2)-dx^2.
\end{equation}
To find the unknown coefficients of the metric (\ref{met}), we need
to solve the Einstein's field equations. This will be done in the
next section.
\\
The covariant components , $\zeta_a=g_{ab}\zeta^b$, of the conformal
vector are
\begin{equation}\label{cov}
\begin{array}{ccc}
\zeta_0=c_0e^{2A},\\
\zeta_1=-e^{A},\\
\zeta_2=-c_2e^{2B},\\
\zeta_3=-c_3e^{2B}.
\end{array}
\end{equation}
It is clear from equations (\ref{zeta^a}) and (\ref{cov}) that  the
obtained conformal vector is non-null conformal vector field,  where
$\zeta^a\zeta_a\neq 0$.

\setcounter{equation}{0}
\section{Solutions of Einstein's field equations  for a perfect fluid}
In this section, we will assume that the space-time under study
admits  a conformal vector field (conformal motion) and then solve
the Einstein's field equations by considering that the matter in
this space-time is represented by a perfect fluid. In general, the
Einstein field equations are of the following form:
\begin{equation}\label{EFEs}
R_{ab}-\frac{1}{2}Rg_{ab}=\kappa T_{ab},
\end{equation}
where $R_{ab}$ is the Ricci tensor, $R$ the Ricci scalar and
$T_{ab}$  the stress energy tensor, which describes the matter field
in the space-time. In Equation (\ref{EFEs}), $\kappa$ is the
coupling constant defined by $\kappa= \frac{8\pi G}{c^4}$, where $G$
is a Newton's gravitational constant and $c$ the speed of light in
vacuum. (For convenience, we assumed that natural units $c=8\pi
G=1$). In the case of a perfect fluid the energy momentum tensor,
$T_{ab}$, is

\begin{equation}\label{EMT}
T_{ab}=(\rho +p)u_au_b-pg_{ab},
\end{equation}
where $p$ is the pressure, $\rho$ the energy density and $u_a$ the
four velocity vector. The contravariant and covariant components of
the 4-velocity
vector, for the space-time (\ref{met}),  can be defined by
$u^a=(e^{-A},0,0,0), u_a=(e^{A},0,0,0)$, and they are verified $g_{ab}u^a u^b=1$.\\
For the line element (\ref{met}), equations (\ref{EFEs}) and
(\ref{EMT}) give the following system of equations
\begin{equation}\label{E1}
3B^{\prime 2}+2B^{\prime\prime}=\rho
\end{equation}
\begin{equation}\label{E2}
B^{\prime 2}+2A^\prime B^\prime=- p
\end{equation}
\begin{equation}\label{E3}
B^{\prime 2}+B^{\prime\prime}+A^\prime B^\prime+A^{\prime
2}+A^{\prime\prime}=- p
\end{equation}
From equations (\ref{E2}) and (\ref{E3}), we have
\begin{equation}\label{E4}
A^{\prime\prime}+B^{\prime\prime}-A^\prime B^\prime+A^{\prime 2}=0.
\end{equation}
 Using the constraint relation given in theorem (2.1), $A=B$,
 in the above equation, we
obtain
\begin{equation}
A^{\prime\prime}=B^{\prime\prime}=0.
\end{equation}
Integrating this equation, we get
\begin{equation}
A=B=ax+b,
\end{equation}
where $a$ and $b$ are constants of integration.\\
As a result, the exact conformal solution of the Einstein's field
equations for a static plane symmetric space-time (\ref{met}) is
given by the following form
\begin{equation}\label{met1}
ds^2=e^{2(ax+b)}dt^2-dx^2-e^{2(ax+b)}(dy^2+ dz^2),
\end{equation}
and $p$ and $\rho$ (the physical variables)  are
\begin{equation}
\rho=-p=3a^2.
\end{equation}
Now, we can conclude that in the case of a perfect fluid, the
assumption of conformal symmetry reproduces the well-known static
plane symmetric solutions (\ref{met})  to give the anti-De Sitter in
the following form
\begin{equation}\label{met11}
ds^2=e^{2(ax+b)}(dt^2-dy^2- dz^2)-dx^2.
\end{equation}

\setcounter{equation}{0}
\section{ Field equations and their solutions for a cosmic strings cloud}
In this section, we will study the gravitational effects for the
space-time (\ref{met}) of a cosmic strings cloud. To do this study
we consider the Einstein's field equations (\ref{EFEs}) in a mixed
form and the stress tensor $T^a_b$ takes the following form
\begin{equation}\label{Xa}
T^a_b=\mu u^au_b -\lambda X^aX_b,
\end{equation}
where $\mu$  and $\lambda$ are  a rest energy density and a string
cloud tension density for a string cloud with particles
attached to it.\\
Here, $u^a$ is a four-velocity vector of particles, as defined
before, and ${\bf{X}}$  a  unit space-like vector representing the
direction of strings orthogonal to  a four-velocity vector. the
vector ${\bf{X}}$ must be taken along any of the three directions
$\frac{\partial}{\partial x}, \frac{\partial}{\partial y},
\frac{\partial}{\partial z}$. For the space-time under
consideration, we choose ${\bf{X}}$ to be parallel to
$\frac{\partial}{\partial x}$, so that $X^0=X^2=X^3=0,$ and $X^1\neq
0$. The components of the  vectors ${\bf{u}}$ and ${\bf{X}}$ satisfy
the following conditions
\begin{equation}
u_au^a=-X_aX^a=1, \quad u_aX^a=0.
\end{equation}
For the space-time (\ref{met}), in a comoving coordinate system, we
get
$$
X^a=(0,1,0,0),
$$
$$
 X_a=(0, -1, 0,0).
$$
If we define the particle density of the configuration  by $\mu_p$,
then the relation between a rest energy density $\mu$ and a string
cloud tension density $\lambda$ is given by
\begin{equation}\label{mulambda}
\mu=\mu_p +\lambda.
\end{equation}
Using equations (\ref{Xa})- (\ref{mulambda}) in Einstein's field
equations (\ref{EFEs}), then for a plane symmetric space-time
(\ref{met}),  we have the following equations
\begin{equation}\label{G11}
B^{\prime 2}+2A^\prime B^\prime=\lambda,
\end{equation}
\begin{equation}\label{G22}
B^{\prime 2}+B^{\prime\prime}+A^\prime B^\prime+A^{\prime
2}+A^{\prime\prime}=0,
\end{equation}
\begin{equation}\label{G00}
3B^{\prime 2}+2B^{\prime\prime}=\mu.
\end{equation}
Before solving  the above  Einstein's field equations, we note that
if we assumed that the direction of the strings is parallel to
$\frac{\partial}{\partial y}$ (or $\frac{\partial}{\partial z}$),
the left-hand side of the equation ( \ref{G11}) equals zero,
therefore either $B=const.$ or $B^\prime +2A^\prime=0$. According to
Theorem 2.1  ($A=B$), the later  gives $A=B=const.$. Therefore, the
space-time under consideration (\ref{met}) becomes flat. So the
direction of the fluid is taken to be in the direction of
$\frac{\partial}{\partial x}$.
\par
Since the  scale factors $A$ and $B$ appearing in the left
hand-sides of equations (\ref{G11})-(\ref{G00}) are functions of $x$
only, then $\lambda$ and $\mu$ must be functions of x only.\\
As we previously indicated in the introduction, we will consider
that the space-time under study admits a conformal motion.
Considering the current case and using Theorem 2.1, the previous
equations (\ref{G11})-(\ref{G00}) reduce to the following equations
\begin{equation}\label{G1}
\lambda=3B^{\prime 2},
\end{equation}
\begin{equation}\label{G2}
3B^{\prime 2}+2B^{\prime\prime}=0,
\end{equation}
\begin{equation}\label{G0}
3B^{\prime 2}+2B^{\prime\prime}=\mu.
\end{equation}
From equations (\ref{G2}) and (\ref{G0}), we get
$$
\mu=0.
$$
Integrating equation (\ref{G2}), using Theorem 2.1, we obtain
\begin{equation}\label{B}
B=\ln(\frac{3}{2}x+c_4)^{\frac{2}{3}}+c_5=A,
\end{equation}
where $c_4$ and $c_5$ are constant of integration.\\
Inserting equation (\ref{B}) into equation (\ref{G1}), we get
(assuming $c_5=0$)
\begin{equation}\label{lambda}
\lambda=\frac{3}{(\frac{3}{2}x+c_4)^2}.
\end{equation}
Since $\mu=\mu_p+\lambda$, then
\begin{equation}\label{mup}
\mu_p=-\frac{3}{(\frac{3}{2}x+c_4)^2}.
\end{equation}
From the equation (\ref{lambda}) and (\ref{mup}) one can see that
the tension density and particle density in the strings decrease as
the fluid moves along the $x$-axis  and  the two densities vanish as
$x\rightarrow\infty$ .\\
 As a result, in  case of a cosmic strings
cloud, the exact conformal solution of the Einstein's field
equations for a static plane symmetric space-time (\ref{met}) is
given by the following form
\begin{equation}\label{met2}
ds^2=(\frac{3}{2}x+c_4)^{\frac{4}{3}}(dt^2--dy^2- dz^2)-dx^2.
\end{equation}
As in the case of perfect fluid the metric (\ref{met}) reduces to
the anti-De Sitter metric.
 \setcounter{equation}{0}
\section{ Orthogonal and parallel  conformal vector fields}
Discussion of conformal vector fields orthogonal or parallel to the
four-speed vector field gives some physical properties of a given
space-time. In this section, we study the two cases of the
space-time (\ref{met}).
\subsection*{conformal vector orthogonal to $u^a$}
In this case
$$
\zeta^au_a=0.
$$
From the definition of the 4-velocity vector, we get
$$
\zeta^0 =0.
$$
Using these results, equations (\ref{1}) - (\ref{10}) are reduced to
the following equations:
\begin{equation}\label{1-}
A^\prime \zeta^1 =\psi(x),
\end{equation}
\begin{equation}\label{2-}
\zeta^1_{,0} =\zeta^2_{,0} =\zeta^3_{,0} =0,
\end{equation}
\begin{equation}\label{5-}
\zeta^1_{,1} =\psi(x),
\end{equation}
\begin{equation}\label{6-}
e^{2B}\zeta^2_{,1}+ \zeta^1_{,2}=0,
\end{equation}
\begin{equation}\label{7-}
e^{2B}\zeta^3_{,1}+ \zeta^1_{,3}=0,
\end{equation}
\begin{equation}\label{8-}
\zeta^2_{,2} +B^\prime\zeta^1=\psi(x),
\end{equation}
\begin{equation}\label{9-}
\zeta^2_{,3}+ \zeta^3_{,2}=0,
\end{equation}
\begin{equation}\label{10-}
\zeta^3_{,3} +B^\prime\zeta^1=\psi(x).
\end{equation}

From equations (\ref{1-}) and (\ref{5-}), we get
\begin{equation}\label{11-}
\zeta^1=\frac{\psi(x)}{A^\prime}, \qquad \zeta^1=\int{\psi(x)dx}
+F(y,z),
\end{equation}
where $F(y,z)$ is an arbitrary function to be determined. Using the
results (\ref{11-})  back into the above equations, we get $F(y,z)$=
constant, without loss of generality, we take it equal zero.
Integrating the result obtained from (\ref{11-}), we get
\begin{equation}\label{12-}
A=\ln(\int{\psi(x)dx}) + c,
\end{equation}
where $c$ is a constant of integration. Using (\ref{12-}) in
equations (\ref{1-})-(\ref{10-}), we have
$$
\zeta^1=\zeta^1(x), \zeta^2 =  c_2, \zeta^3=c_3,
$$
where $c_2$ and $c_3$ are constants of integration, which will be
taken as zeros. We have also
\begin{equation}\label{13-}
A^\prime = B^\prime \quad \Rightarrow \quad A=B.
\end{equation}
Consequently,  the conformal vector field  orthogonal to the
4-velocity vector is
\begin{equation}\label{14-}
\zeta_\perp=\frac{\psi(x)}{A^\prime}\frac{\partial}{\partial x}.
\end{equation}
To verify that the resulting vector is a proper conformal vector,
that is, we prove that the conformal factor are function in the
coordinate $x$, we use the above results in Einstein's field
equations.\\
{\bf{Case I}}: Perfect fluid case\\
In this case,  using equations (\ref{12-}) and (\ref{13-})  in
Einstein's field equations (\ref{E1})- (\ref{E3}), we get the
following equations
\begin{equation}\label{E1-}
\frac{\psi^2(x)}{(\int{\psi(x)dx})^2}+\frac{2\psi^\prime(x)}{\int{\psi(x)dx}}=\rho
\end{equation}
\begin{equation}\label{E2-}
\frac{3\psi^2(x)}{(\int{\psi(x)dx})^2}=- p
\end{equation}
\begin{equation}\label{E3-}
\frac{3\psi^2(x)}{(\int{\psi(x)dx})^2}+2\frac{\psi^\prime(x)\int{\psi(x)dx}-\psi^2(x)}{(\int{\psi(x)dx})^2}=-
p
\end{equation}
From equations (\ref{E2-}) and (\ref{E3-}), we have
\begin{equation}\label{E4-}
\psi^\prime(x)\int{\psi(x)dx}-\psi^2(x)=0.
\end{equation}
The solution of this equation is
\begin{equation}
\psi(x)=e^{mx},
\end{equation}
where $m$ is a constant of integration. Using the above equation in
(\ref{12-}), the scale factor $A$ can be written as follows
\begin{equation}
A=ax +b=B.
\end{equation}
Now the previous discussion can be summarized in the following:
\begin{prop}
All perfect fluid solutions described by the metric ansatz
(\ref{met}) admit a conformal Killing vector field,
$\zeta_\perp=\frac{e^{mx}}{a}\frac{\partial}{\partial x}$,
orthogonal to the 4-velocity vector, ${\bf{u}}$, if the scale
factors are
$$
A=B=ax+b,
$$
and the conformal factor is
$$
\psi(x)=e^{mx}.
$$
\end{prop}
According to the above proposition, the dynamical variables are
$$
\rho=-p=3m^2.
$$
{\bf{Case II}}: Cosmic string cloud case\\
Inserting (\ref{12-}) and (\ref{13-}) into (\ref{G1}) - (\ref{G0}),
we get $\mu =0$ and
$$
3A^{\prime 2}+2A^{\prime\prime}=0
$$
Using (\ref{12-}) in the above equation and integrating the obtain
results,  we obtain the conformal factor and scale factors,
respectively, as follows
$$
\psi(x)= (\frac{2q}{3x})^{\frac{1}{3}},
$$
$$
A=\ln(mx^{\frac{2}{3}})+n=B,
$$
where $q$ and $n$ are constants of integration and $m=(\frac{9q}{4})^{\frac{1}{3}}$.\\
The previous discussion can be summarized as the following
\begin{prop}
All cosmic string cloud solutions described by the metric
(\ref{met}) admit a conformal Killing vector field,
$\zeta_\perp=\frac{x^2}{4q}\frac{\partial}{\partial x}$, orthogonal
to the 4-velocity vector ${\bf{u}}$ if the  conformal factor and
scale factors are, respectively
$$
\psi(x)= (\frac{2q}{3x})^{\frac{1}{3}}, \quad
A=B=\ln(mx^{\frac{2}{3}})+n.
$$
\end{prop}

\subsection*{conformal vector parallel to $u^a$ ($\zeta_\parallel$)}
In this case
$$
\zeta^a \propto u^a,
$$
then
$$
\zeta^a =\ell u^a,
$$
 where $\ell$  is a constant of
proportionality. From the definition of $u^a$ for the space-time
(\ref{met}), the above equation gives
$$
\zeta^1=\zeta^2=\zeta^3=0.
$$
Then the conformal equations (\ref{1}) - (\ref{10}) reduce to the
following
\begin{equation}\label{parallel}
\zeta^0_{,0}=\psi(x),
\end{equation}
$$
\zeta^0_{,1}=\zeta^0_{,2}=\zeta^0_{,3}=0.
$$
Then the parallel conformal vector field is
$$
\zeta_\parallel = \zeta_\parallel(t).
$$
Therefore, equation (\ref{parallel}) gives $\psi=$constant or
$\psi=0$, that is, $\zeta_\parallel=0$ or equal constant. So the
space-time (\ref{met}) does not admit a conformal vector field
parallel to the 4-velocity vector.

\setcounter{equation}{0}
\section{ Physical properties}
In the previous three sections, we discussed  a static plane
symmetric space-time (\ref{met}) and attempted to obtain exact
solutions to the Einstein field equations (\ref{EFEs}). To do this,
in addition to considering the space-time under study admits
conformal symmetry, we assumed that the matter is represented by an
a perfect fluid, as in Section 3, and solved the field equations for
a cosmic string cloud in Section 4. To discuss the physical behavior
of the obtained conformal solutions given by the metrics
(\ref{met1}) and (\ref{met2}), We need to find the following
physical and kinematical parameters of the model which are very
important to give us a deeper insight into the properties of the
cosmology.\\
 For the metric (\ref{met1}), we find the following parameters:
\begin{enumerate}
\item The non-vanishing component of the 4-acceleration is
$$
\dot{u}_1=const. =-a
$$
\item The non-vanishing component of the rotation is
$$
\omega_{01}=-\omega_{10}=ae^{ax+b}.
$$
\end{enumerate}
For the metric (\ref{met2}), the above components are
$$
\dot{u}_1=-\frac{1}{2(\frac{3}{2}x+c_4)},
$$
$$
\omega_{01}=-\omega_{10}=
\frac{1}{2(\frac{3}{2}x+c_4)^{\frac{2}{3}}}
$$

\section{Discussion and conclusion}
One of the most common attempts to obtain exact solutions to
Einstein's field equations is to assume symmetries in space-time..
These symmetries are defined by operating
 the Lie derivative of the considered tensor, such as,  $g_{ab},
\Gamma^{a}_{bc}, T_{ab}, R_{ab}$, ... etc, with respect to  space,
time, or null vector. The resulting geometric objects created  by
these operators are  tensors with the same index or zero.\\
 This
work is devoted to studying one of these symmetries, in particular
conformal symmetry, of a plane static symmetric model in the
framework of general relativity.
 We focused on this type of
symmetries because a space-time admitting it preserves the causal
character of space-time, So it is in an important physical form one.
For a static plane symmetric space-time, we  solved the conformal
equations and obtained the conformal vector field that the
space-time admits. Furthermore, solving these equations helped us to
obtain a relationships between the metric coefficients. We have used
these relationship to simplifying Einstein's field equations and got
the energy density ($\rho$) and pressure ($p$) (dynamical
variables), which depend on the coordinates $x$. We obtained new
exact  solutions of the Einstein's field equations for static plane
symmetric space-times by considering that they admit conformal
symmetry. In the case of a perfect fluid, the resulting solutions
have negative  pressure, which represents a possible example of a
dark energy star, and the energy equation of state is $\rho +p=0$..
Moreover, we have shown that these solutions reduce to the
well-known anti-De Sitter space-times, when the energy-momentum
tensor is represented by a perfect fluid or cosmic strings cloud. In
the case of a cosmic string cloud, we found that the tension density
and particle density decrease as the fluid moves
along the direction of the strings, and then vanish at infinity.\\
For the solutions obtained, all coefficients of the metric are well
defined so there is no singularity present. They have acceleration,
rotation, shear-free, vanishing expansion, and rotation. We have
discussed the orthogonal and parallel conformal vector fields and
obtained that the space-time under consideration admits a conformal
vector field orthogonal to the four-velocity vector but does not
admits a vector parallel to the four-velocity vector.


\end{document}